\documentclass[aps,prl,citeautoscript,twocolumn,showpacs,superscriptaddress]{revtex4-1}  

\usepackage{graphicx}  
\usepackage{dcolumn}   
\usepackage{bm}        
\usepackage{amssymb}   
\usepackage{amsmath}   
\usepackage{algorithm}
\usepackage{algpseudocode}
\usepackage[dvipsnames]{xcolor}
\usepackage{soul}
\usepackage{multirow}
\usepackage{tabularx}
\usepackage{array}
\usepackage{makecell}
\usepackage{chemformula}

\newcommand{\xcdnn}{XCNN}

\begin{document}

\title{Learning the exchange-correlation functional from nature with fully differentiable density functional theory
}

\author{M. F. Kasim}
\email{muhammad.kasim@physics.ox.ac.uk}
\affiliation{Department of Physics, Clarendon Laboratory, University of Oxford, Parks Road, Oxford OX1 3PU, UK}
\author{S. M. Vinko}
\email{sam.vinko@physics.ox.ac.uk}
\affiliation{Department of Physics, Clarendon Laboratory, University of Oxford, Parks Road, Oxford OX1 3PU, UK}
\affiliation{Central Laser Facility, STFC Rutherford Appleton Laboratory, Didcot OX11 0QX, UK}

\date{\today}

\begin{abstract}

Improving the predictive capability of molecular properties in {\it ab initio} simulations is essential for advanced material discovery.
Despite recent progress making use of machine learning, utilizing deep neural networks to improve quantum chemistry modelling remains severely limited by the scarcity and heterogeneity of appropriate experimental data.
Here we show how training a neural network to replace the exchange-correlation functional within a fully-differentiable three-dimensional Kohn-Sham density functional theory (DFT) framework can greatly improve simulation accuracy.
Using only eight experimental data points on diatomic molecules, our trained exchange-correlation networks enable improved prediction accuracy of atomization energies across a collection of 104 molecules containing new bonds and atoms that are not present in the training dataset.


\end{abstract}


\maketitle

Fast and accurate predictive models of atomic and molecular properties are key in advancing novel material discovery. With the rapid development of machine learning (ML) techniques, considerable effort has been dedicated to accelerating predictive simulations, enabling the efficient exploration of configuration space. Much of this work focuses on end-to-end ML approaches~\cite{gilmer2017-neural-message-passing}, where the models are trained to predict some specific molecular or material property using tens to hundreds of thousands of data points, generated via simulations~\cite{schutt2018schnet}. Once trained, ML models have the advantage of accurately predicting the simulation outputs much faster than the original simulations~\cite{smith2017ani-1}. The level of accuracy achieved by ML models with respect to the simulated data can be very high, and may even reach chemical accuracy on select systems outside the original training dataset~\cite{faber2017prediction-ml-better}. Although these machine learning approaches show considerable promise, they rely heavily on the availability of large datasets for training.

While there has been a concerted effort in accelerating predictive models, less attention has been dedicated to using machine learning to increase the accuracy of the model itself with respect to the experimental data.
In particular, as ML models typically rely on some underlying simulation for the training data, the trained models can never exceed the simulations in terms of accuracy.
However, training ML models using experimental data is challenging as the available data is highly heterogeneous as well as limited, making it difficult to train accurate ML models tailored to reproducing a specific system or physical property.
This limitation also poses a great challenge to the generalization capability of ML models to systems outside the training data.

One emerging solution to this problems is to use machine learning to build models which learn the exchange-correlation (xc) functional within the framework of Density Functional Theory (DFT)~\cite{snyder2012ml-xc,lei2019design-xc, nagai2020completing,dick2020mlxc, li2021kohn-sham-regularizer,chen2020deepks}. Here, the Kohn-Sham scheme (KS-DFT)~\cite{hohenberg1964inhomogeneous,kohn1965self-ks} can be used to calculate various ground-state properties of molecules and materials, the accuracy of which largely depends on the quality of the employed xc functional.
Work along these lines in the one-dimensional case show promise~\cite{snyder2012ml-xc, li2021kohn-sham-regularizer}, but are severely limited by the lack of availability of experimental data for one-dimensional systems.
An alternative approach is to learn the xc functional using a dataset constructed from electron density profiles obtained from coupled-cluster (CCSD) simulations~\cite{vcivzek1966correlation-ccsd}, and the xc energy densities obtained from an inverse Kohn-Sham scheme \cite{kanungo2019exact-inverse-ks}. However, this method is fundamentally limited to, at most, match the accuracy of the original simulations.
Recently, Nagai \textit{et al.}~\cite{nagai2020completing} reported on successfully representing the xc functional as a neural network, trained to fit 2 different properties of 3 molecules using the three-dimensional KS-DFT calculation in PySCF~\cite{sun2018pyscf} via gradient-less optimization. While their results show promise in predicting various material properties, gradient-less methods in training neural networks typically suffer from bad convergence~\cite{maheswaranathan2019guided-es} and are too computationally intensive to scale to larger and more complex neural networks.

In this context, we present here a new machine learning approach to learn the xc functional using heterogeneous experimental data. We achieve this by incorporating a deep neural network representing the xc functional in a fully differentiable three-dimensional KS-DFT calculation. In our method, the xc functional is represented by a network that takes the three-dimensional electron density profile as its input and produces the xc energy density as its output. Notably, using a xc neural network (\xcdnn{}) in a fully differentiable KS-DFT calculation gives us a framework to train the xc functional using any available experimental properties, however heterogeneous, which can be simulated via standard KS-DFT. Moreover, as the neural network is independent of the specific physical system (atom type, structure, number of electrons, {\it etc}.), a well-trained \xcdnn{} should be generalizable to a wide range of systems, including those outside the pool of training data.
In this paper, we demonstrate this approach using a dataset comprising atoms and molecules from the first few rows of the periodic table, but the results are readily extendable to more complex structures.


A critical part of realizing our method is writing a KS-DFT code in a fully differentiable manner.
This is achieved here by writing the KS-DFT code using an automatic differentiation library, PyTorch \cite{neurips2019-pytorch}, and by providing separately the gradients of all the KS-DFT components which are not available in PyTorch.
The KS-DFT calculation involves self-consistent iterations of solving the Kohn-Sham equations,
\begin{align}
    \label{eq:eigdecomp}
    H[n](\mathbf{r}) \phi_i(\mathbf{r}) &= \varepsilon_i \phi_i(\mathbf{r}), \\
    \label{eq:density-profile}
    n(\mathbf{r}) &= \sum_i f_i |\phi_i(\mathbf{r})|^2,
\end{align}
where $n(\mathbf{r})$ is the electron density as a function of a 3D coordinate ($\mathbf{r}$), $f_i$ is the occupation number of the $i$-th orbital ($\phi_i$), $\varepsilon_i$ is the Kohn-Sham orbital energy, and $H[n](\mathbf{r})$ is the Hamiltonian operator, a functional of the electron density profile $n$ and a function of the 3D coordinate.
The Hamiltonian operator is given in atomic units as $H[n](\mathbf{r}) = -\nabla^2/2 + v(\mathbf{r}) + v_H[n](\mathbf{r}) + v_{xc}[n](\mathbf{r})$ with $-\nabla^2/2$ representing the kinetic operator, $v(\mathbf{r})$ the external potential operator including ionic Coulomb potentials, $v_H[n](\mathbf{r})$ the Hartree potential operator, and $v_{xc}[n](\mathbf{r})$ the xc potential operator.
The xc potential operator is defined as the functional derivative of the xc energy with respect to the density, i.e. $v_{xc}[n](\mathbf{r}) = \delta E_{xc}[n] / \delta n(\mathbf{r})$.

In this work, the orbitals in the above equations are represented using contracted Gaussian-type orbital basis sets~\cite{pritchard2019new-bse}, $b_j(\mathbf{r})$ for $j=\{1...n_b\}$ where $n_b$ is the number of basis elements.
As the basis set is non-orthogonal, Eq.~(\ref{eq:eigdecomp}) becomes the Roothaan's equation \cite{roothaan1951new},
\begin{equation}
    \label{eq:roothaan}
    \mathbf{F}[n]\mathbf{p}_i = \varepsilon_i \mathbf{Sp}_i,
\end{equation}
where $\mathbf{p}_i$ is the basis coefficient for the $i$-th orbital, $\mathbf{S}$ is the overlap matrix with elements $S_{rc}=\int b_r(\mathbf{r}) b_c^*(\mathbf{r})\mathrm{d}\mathbf{r}$, and $\mathbf{F}[n]$ is the Fock matrix as a functional of the electron density profile, $n$, with elements $F_{rc}=\int b_r(\mathbf{r}) H[n](\mathbf{r}) b_c^*(\mathbf{r})\mathrm{d}\mathbf{r}$.
Unless specified otherwise, all calculations involved in this paper use 6-311++G(3pd,3df) basis sets \cite{clark1983efficient-pople-set}.

\begin{figure}
    \centering
    \includegraphics[width=\linewidth]{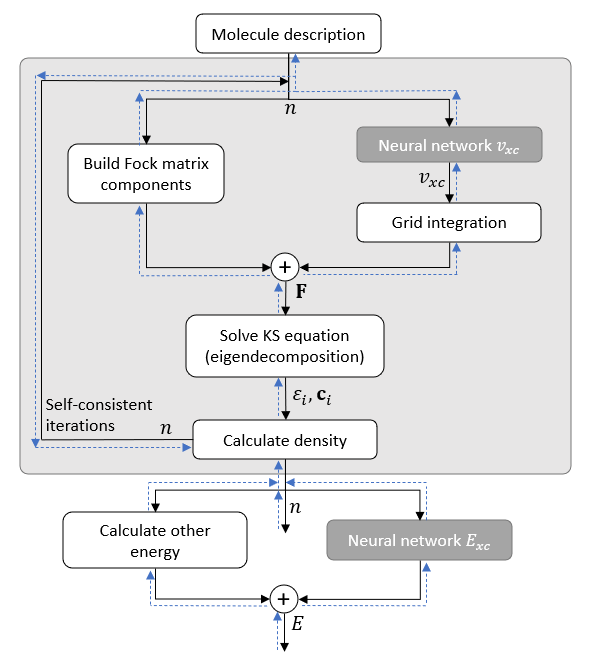}
    \caption{Schematics of the xc neural network and differentiable KS-DFT code.
    The trainable neural network is illustrated with darker gray boxes.
    The solid dark lines show the forward calculation flow while the dashed blue lines show the gradient backpropagation flow.
    }
    \label{fig:schematics}
\end{figure}

Equation~(\ref{eq:roothaan}) can be solved by performing an eigendecomposition of the Fock matrix to obtain the orbitals from the electron density profile.
The ionic Coulomb potential, Hartree potential, and the kinetic energy operators in the Fock matrix are constructed by calculating the Gaussian integrals with libcint~\cite{sun2015libcint}.
The contribution from the xc potential operator in the Fock matrix is constructed by integrating it with the basis using the SG-3 integration grid~\cite{dasgupta2017standard-sg3}, with Becke scheme to combine multi-atomic grids~\cite{becke1988multicenter}.
The KS-DFT calculation proceeds by solving Eq.~(\ref{eq:roothaan}) and computing Eq.~(\ref{eq:density-profile}) iteratively until convergence or self-consistency is reached. The schematics of our differentiable KS-DFT code is shown in Fig.~\ref{fig:schematics}.

Although modern automatic differentiation libraries provide many operations that enable an automatic propagation of gradient calculations, there are many gradient operations that have to be implemented specifically for this project.
An important example is the self-consistent cycle iteration.
Previous approaches employed a fixed number of linear mixing iterations to achieve self-consistency, and let the automatic differentiation library propagate the gradient through the chain of iterations~\cite{tamayo2018automatic-diffiqult, li2021kohn-sham-regularizer}.
In contrast, we implement the gradient propagation of the self-consistency cycle using an implicit gradient calculation (see supplementary materials), allowing us to deploy a better self-consistent algorithm, achieving faster and more robust convergence than linear mixing. We implemented the gradient calculation of the self-consistency cycle in xitorch~\cite{kasim2020xitorch}, a publicly available computational library written specifically for this project.

Another example is the gradient calculation of the eigendecomposition with (near-)degenerate eigenvalues.
This often produces numerical instabilities in the gradient calculation.
To solve this problem, we follow the calculation by Kasim~\cite{kasim2020derivatives-degen} to obtain a numerically stable gradient calculation in cases of (near-)degeneracy, which is also implemented in xitorch.
Other parts in KS-DFT that require specific gradient implementations are the calculation of the xc energy, the Gaussian basis evaluation, and the Gaussian integrals. The details of the gradient calculations in these cases can be found in the supplementary materials.


With the fully differentiable KS-DFT code to hand, the xc energy represented by a deep neural network can be trained efficiently.
For this work we will test two different types of xc deep neural networks (\xcdnn{}) based on two rungs of Jacob's ladder -- the Local Density Approximation (LDA) and the Generalized Gradient Approximation (GGA).
These xc energies can be written as:
\begin{align}
    E_{\rm nnLDA}[n] &= \alpha E_{\rm LDA(xc)}[n] + \beta \int n(\mathbf{r}) f(n, \xi)\ \mathrm{d}\mathbf{r}\\
    E_{\rm nnPBE}[n] &= \alpha E_{\rm PBE(xc)}[n] + \beta \int n(\mathbf{r}) f(n, \xi, s)\ \mathrm{d}\mathbf{r},
\end{align}
where $E_{\rm LDA(xc)}$ is the LDA exchange~\cite{kohn1965self-ks} and correlation energy~\cite{perdew1992accurate-pw92}, and $E_{\rm PBE(xc)}$ is the PBE exchange-correlation energy~\cite{perdew1996generalized-gga-pbe}.
Here $\alpha$ and $\beta$ are two tunable parameters, and $f(\cdot)$ is the trainable neural network. The LDA and PBE energies are calculated using Libxc~\cite{lehtola2018recent} and wrapped with PyTorch to provide the required gradient calculations.

The \xcdnn{}s take inputs of the local electron density $n$, the relative spin polarization $\xi = (n_\uparrow - n_\downarrow) / n$, and the normalized density gradient $s = |\nabla n| / (24\pi^2 n^4)^{1/3}$~\cite{perdew1996generalized-gga-pbe}.
The neural network is an ordinary feed-forward neural network with 3 hidden layers consist of 32 elements each with a softplus~\cite{nair2010rectified-relu-softplus} activation function to ensure infinite differentiability of the neural network.
To avoid non-convergence of the self-consistent iterations during the training, the parameters $\alpha$ and $\beta$ are initialized to have values of 1 and 0 respectively.

 
For the \xcdnn{} training, we compiled a small dataset consisting of experimental atomization energies (AE) from the NIST CCCBDB database~\cite{NIST_CCCBDB}, atomic ionization potentials (IP) from the NIST ASD database~\cite{NIST_ASD}, and calculated density profiles using a CCSD calculation~\cite{vcivzek1966correlation-ccsd}.
The calculated density profiles are used as a regularization to ensure that the learned xc does not produce electron densities that are too far from CCSD predictions, a concern discussed by Medvedev {\it et al.}~\cite{medvedev2017density-deviate}.

The dataset is split into 2 groups: training and validation.
The training dataset is used directly to update the parameters of the neural network via the gradient of a loss function (see supplementary materials for details on the loss function).
After the training is finished, we selected a model checkpoint during the training that gives the lowest loss function for the validation dataset.

The complete list of atoms and molecules used in the training and validation datasets is shown in Table \ref{tab:tvdset}.
Importantly, the size of the dataset used in this work is orders of magnitude smaller than datasets used in most ML-related work present in the literature on predicting molecular properties~\cite{ramakrishnan2014quantum-qm9, chmiela2017machine-md17, brockherde2017bypassing-ks}.
We note that some atoms are only present in the validation set and not in the training set, to include checkpoints that are able to generalize well to new types of atoms outside the training set.

\newcommand{\dsetcolwidth}{0.25\linewidth}
\newcommand{\dsetsmallspacing}{-2.0ex}
\newcommand{\dsetsmallspacingafter}{-2.2ex}
\begin{table}[b]
    \caption{Atoms and molecules presented in the training and validation datasets.}
    \label{tab:tvdset}
    \centering
    \begin{ruledtabular}
    \begin{tabular}{lccc}
        Type &
        \parbox[t]{\dsetcolwidth}{Atomization energy} &
        \parbox[t]{\dsetcolwidth}{Density profile} &
        \parbox[t]{\dsetcolwidth}{Ionization potential} \\ [\dsetsmallspacingafter] \\
        
        \hline \\ [\dsetsmallspacing]
        \multirow{1}{*}{Training} & 
        \parbox[t]{\dsetcolwidth}{H$_2$, LiH, \\ O$_2$, CO} & 
        \parbox[t]{\dsetcolwidth}{H, Li, Ne, H$_2$, Li$_2$, LiH, B$_2$, O$_2$, CO} &
        \parbox[t]{\dsetcolwidth}{O, Ne} \\ [\dsetsmallspacingafter] \\
        
        
        \hline \\ [\dsetsmallspacing]
        \multirow{1}{*}{Validation} &
        \parbox[t]{\dsetcolwidth}{N$_2$, NO, \\ F$_2$, HF} & 
        \parbox[t]{\dsetcolwidth}{He, Be, N, N$_2$, F$_2$, HF} &
        \parbox[t]{\dsetcolwidth}{N, F} \\ [\dsetsmallspacingafter] \\
        
        
    \end{tabular}
    \end{ruledtabular}
\end{table}


\newcommand{\bestresult}[1]{\textbf{\underline{#1}}}
\newcommand{\bestgroup}[1]{\textbf{#1}}
\begin{table*}
    \caption{Mean absolute error (MAE) in kcal/mol of the atomization energy and ionization potential for atoms and molecules in the test dataset.
    The column ``IP 18'' represents the deviation in ionization potential for atoms H-Ar.
    Column ``AE 104'' is the MAE of atomization energy of a collection of 104 molecules from ref.~\cite{curtiss1997assessment-g2}.
    Column ``DP 99'' is the difference of the density profile (in $\times 10^{-3}$ Bohr$^{-3}$) for the 104 molecules, excluding 5 molecules that have multiple possible density profiles.
    The following 4 columns present the atomization energies of subsets of molecules from the full 104-molecule set: ``AE 16 HC'' for 16 hydrocarbons, ``AE 25 subs HC'' for 25 substituted hydrocarbons, ``AE 34 others-1'' for 34 non-hydrocarbon molecules containing only first and second row atoms, and ``AE 30 others-2'' for 30 non-hydrocarbon molecules containing at least one third row atom.
    The suffix ``-IP'' in the \xcdnn{} calculation indicates that the xc neural networks were trained with the ionization potential dataset.
    The bolded values are the best MAE in the respective column for each group of approximations.}
    \label{tab:results}
    \begin{ruledtabular}
    \begin{tabular}{lrrr|rrrr}
        Calculation & IP 18 & AE 104 & DP 99 & AE 16 HC & AE 25 subs HC & AE 33 others-1 & AE 30 others-2 \\
        \hline
        \multicolumn{2}{l}{\textit{Local density approximations (LDA)}} & & & & \\
        LDA (exchange only) & 24.6 & 28.2 & 25.7 & 48.7 & 29.0 & 25.4 & 19.8 \\
        LDA (PW92~\cite{perdew1992accurate-pw92}) & \bestgroup{6.9} & 70.5 & 23.3 &  97.0 & 101.1 & 58.8 & 43.7 \\
        \xcdnn{}-LDA & 50.8 & \bestgroup{15.5} & 10.9 & \bestgroup{22.7} & \bestgroup{19.4} & \bestgroup{7.8} & \bestgroup{16.6} \\
        \xcdnn{}-LDA-IP & 15.2 & 18.5 & \bestgroup{9.8} & 25.4 & 21.8 & 8.4 & 22.9 \\
        \hline
        \multicolumn{2}{l}{\textit{Generalized gradient approximations (GGA)}} & & & & \\
        PBE \cite{perdew1996generalized-gga-pbe} & \bestgroup{3.6} & 16.5 & 2.6 & 15.1 & 23.2 & 18.0 & 9.8 \\
        \xcdnn{}-PBE & 10.7 & \bestgroup{7.4} & \bestgroup{2.4} & \bestgroup{5.4} & \bestgroup{8.4} & \bestgroup{6.5} & \bestgroup{8.5} \\
        \xcdnn{}-PBE-IP & 4.1 & 8.1 & \bestgroup{2.4} & 6.8 & 9.6 & 6.7 & 8.6 \\
        \hline
        \multicolumn{2}{l}{\textit{Other approximations}} & & & & \\
        SCAN \cite{sun2015strongly-scan} & 3.7 & 5.0 & 1.1 & 3.8 & 8.2 & 4.0 & 4.6 \\
        CCSD (basis: cc-pvqz \cite{dunning1989gaussian-ccpvqz}) & 2.0 & 12.1 & \bestgroup{0.0} & 11.1 & 17.7 & 11.2 & 9.0 \\
        CCSD(T) (basis: cc-pvqz) & \bestgroup{1.3} & \bestgroup{3.5} & N/A & \bestgroup{2.2} & \bestgroup{5.6} & \bestgroup{2.5} & \bestgroup{3.5} \\
    \end{tabular}
    \end{ruledtabular}
\end{table*}

To test the performance of the trained \xcdnn{}s, we prepared a test dataset consisting of the atomization energies of 104 molecules from ref.~\cite{curtiss1997assessment-g2}, and the ionization potentials of atoms H-Ar.
The experimental geometric data as well as the atomization energies for the 104 molecules were obtained from the NIST CCCBDB database~\cite{NIST_CCCBDB}, while the ionization potentials are obtained from the NIST ASD database~\cite{NIST_ASD} (see supplementary materials for the details in compiling the datasets).
There are actually 146 molecules in ref.~\cite{curtiss1997assessment-g2}, however, only 104 molecules have experimental data in CCCBDB, so we limit our dataset to those.
While each molecule in the training and validation datasets only contain 2 atoms from the first two rows of the periodic table (i.e. H-Ne), molecules in the test dataset contain 2-14 atoms from the first three rows of the periodic table (i.e. H-Ar).
A complete list of all the molecules used in the test dataset can be found in the supplementary materials.

Our first experiment is done by training the \xcdnn{} without the ionization potential dataset, i.e., using only atomization energies and density profile data.
The results are presented in Table~\ref{tab:results}.
We see that in this case both \xcdnn{}-LDA and \xcdnn{}-PBE provide significant improvement over their bases, i.e. LDA and PBE, respectively.
In terms of average atomization energy prediction errors across the 104 test molecules, \xcdnn{}-LDA achieves more than 4 times lower error than LDA (PW92) and \xcdnn{}-PBE achieves more than 2 times lower error than PBE.
Importantly, although the training and validation sets contain atomization energies only for 8 diatomic molecules, both \xcdnn{}s provide improvements on atomization energy predictions of the 104-molecule test dataset, which includes molecules with up to 14 atoms.

\begin{figure}[t]
    \centering
    \includegraphics[width=\linewidth]{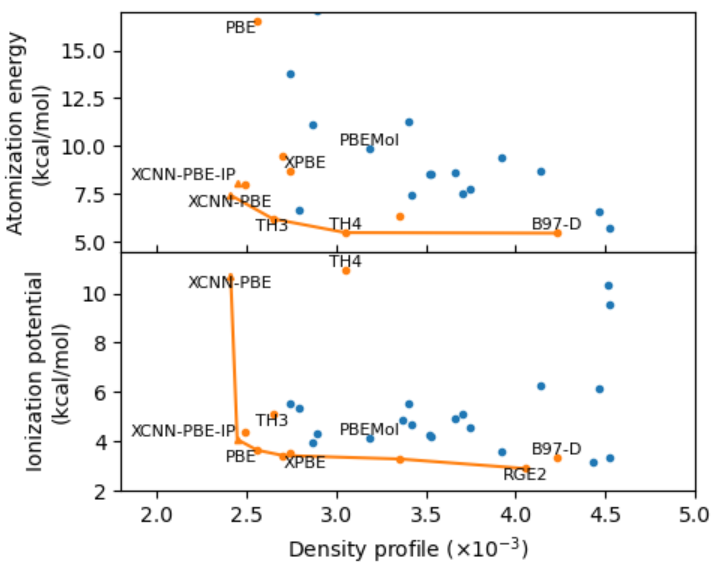}
    \caption{\label{fig:pareto}
    Objective plots for GGA functionals.
    The orange data points are the functionals that lie on the Pareto front for the 3 objectives (atomization energy, density profile, ionization energy).
    The orange line shows the Pareto front on 2 objectives in the respective plots.
    XCNN-PBE and XCNN-PBE-IP are shown by the orange triangle which also lies on the Pareto front.
    The full list can be found in the supplementary materials.
    Only functionals with deviations within the range are shown for clarity.
    }
\end{figure}

\begin{figure}[t]
    \centering
    \includegraphics[width=\linewidth]{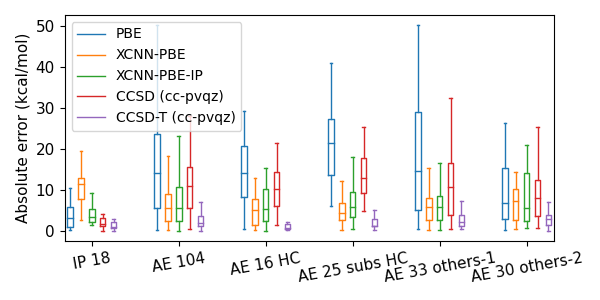}
    \caption{
    Statistical comparison of PBE, XCNN-PBE, CCSD, and CCSD(T) for each group.
    The line in the box represents the median of the absolute error.
    For clarity, outliers are not shown.}
    \label{fig:top-performer}
\end{figure}

By looking into the subsets of the atomization energy dataset, both \xcdnn{}-LDA and \xcdnn{}-PBE also achieve better atomization energy predictions than their bases on all of the subsets.
The substantial improvements seen on the hydrocarbon and substituted hydrocarbon molecules show the generalization capability of the technique to molecules with bonds not present in neither the training nor the validation dataset (e.g., C--H, C--C, C=C).
Moreover, both \xcdnn{}s can also predict better atomization energy on the ``AE-30 others-2'' subset containing third-row atoms that are not present in the training and validation datasets.
The improvement of atomization energy predictions on unseen atoms and bonds demonstrate a very promising generalization capability of the machine learning models outside the training distribution.

Despite these encouraging results on the atomization energies, both \xcdnn{}s lead to significantly worse predictions of atomic ionization potentials compared with their bases.
Therefore, for the next experiment we trained the \xcdnn{} by adding ionization potential data into the training and validation datasets, as shown in Tab.~\ref{tab:tvdset}.
The xc neural networks trained with this additional data are called ``\xcdnn{}-LDA-IP'' and ``\xcdnn{}-PBE-IP''.
Adding ionization potential data to the training and validation datasets improves the performance of \xcdnn{}s on the ionization potential at the cost of slightly increasing the error on the atomization energy predictions.
Although the ionization potential predictions of \xcdnn{}-IPs do not have lower error compared to their bases, they produce considerably better predictions on the atomization energies of the tested molecules than their bases.

When compared to other 48 GGAs on XCNN-PBE and XCNN-PBE-IP give the lowest errors on density profiles deviation.
Moreover, when comparing on atomization energy, density profile, and ionization potential, both XCNN-PBEs lie on the Pareto front as can be seen in Figure \ref{fig:pareto}.
This means that there is no other GGAs that can provide improvement on all aspects considered in this paper.
Detailed results on GGAs comparison can be seen in the supplementary materials.

Another interesting result is that even though CCSD simulation is used in the training and validation datasets, the trained \xcdnn{}-PBE can achieve considerably better predictions on atomization energies than CCSD.
A detailed comparison of our \xcdnn{}-PBE results with CCSD calculations is presented in Fig.~\ref{fig:top-performer}.
We see that the results from \xcdnn{}-PBE are in most cases better than calculations based on CCSD. This result highlights a key advantage of our method over other ML-based methods that rely fully on simulation data: as our method utilizes heterogeneous experimental data in the training, it enables us to learn xc functionals that can exceed in accuracy even the simulations used to train them.

Although the trained \xcdnn{}s cannot make better predictions than higher level of approximations, e.g. SCAN~\cite{sun2015strongly-scan} and CCSD(T), the presented method shows that significant improvement can be made within the same level of approximation.
The level of accuracy can be improved in the future work by using a higher level of approximation, such as moving to the higher rungs of Jacob's ladder or incorporating more global density information with neural networks in calculating the xc energy.
While perfectly viable within our framework, such investigations remain outside the scope of the current paper.


We presented a novel approach to train machine-learned xc functionals within the framework of fully-differential Kohn-Sham DFT using experimental data.
Our xc neural networks, trained on a few atoms and diatomic molecules, are able to improve the prediction of atomization and ionization energies across a set of 104 molecules, including larger molecules containing atoms and bonds not present in the training dataset.
We have limited our study to a minimal training dataset and to a relatively small neural network, but our results are readily extendable to substantially larger systems, bigger experimental datasets, additional heterogeneous physics quantities, and to the use of more sophisticated neural networks.
This demonstration of experimental-data-driven xc functional learning, embedded within a differentiable DFT simulator, shows great promise to advance computational material discovery.
The codes can be found in \cite{kasim2020-xitorch, kasim2021-dqc, kasim2021-xcnn}.

\begin{acknowledgments}
We would like to thank Kyle Bystrom for a useful suggestion on compiling experimental data.
M.F.K. and S.M.V. acknowledge support from the UK EPSRC grant EP/P015794/1 and the Royal Society.
S.M.V. is a Royal Society University Research Fellow.
The authors declare no conflict of interest.
\end{acknowledgments}


\end{document}


\title{Supplementary materials:\\
Learning the exchange-correlation functional from nature with fully differentiable density functional theory
}

\author{M. F. Kasim}
\email{muhammad.kasim@physics.ox.ac.uk}
\affiliation{Department of Physics, Clarendon Laboratory, University of Oxford, Parks Road, Oxford OX1 3PU, UK}
\author{S. M. Vinko}
\email{sam.vinko@physics.ox.ac.uk}
\affiliation{Department of Physics, Clarendon Laboratory, University of Oxford, Parks Road, Oxford OX1 3PU, UK}
\affiliation{Central Laser Facility, STFC Rutherford Appleton Laboratory, Didcot OX11 0QX, UK}

\date{\today}

\maketitle

\section{Gradient of KS-DFT components}

\subsection{Self-consistent iteration cycle}

The self-consistent iteration cycle in a KS-DFT calculation can be written mathematically as
\begin{equation}
    \mathbf{y}(\theta) = \mathbf{f}(\mathbf{y}, \theta),
\end{equation}
where $\mathbf{y}$ are the parameters to be found self-consistently, and $\mathbf{f}(\cdot)$ is the function that needs to be satisfied which may depend additionally on other parameters, here denoted by $\theta$.
The self-consistent iteration takes the parameters $\theta$ and an initial guess $\mathbf{y_0}$, and produces the self-consistent parameters $\mathbf{y}$ that satisfy the equation above.
As the self-consistent $\mathbf{y}$ does not really depend on the initial guess, it is only a function of the parameters $\theta$.
In our case, the self-consistent parameters are the elements of the Fock matrix.

The gradient of a loss function $\mathcal{L}$, with respect to the parameters $\theta$, given the gradient with respect to the self-consistent parameters $\partial\mathcal{L}/\partial\mathbf{y}$, can be expressed as 
\begin{align}
    \frac{\partial \mathcal{L}}{\partial \theta} &= \frac{\partial \mathcal{L}}{\partial \mathbf{y}}\left(\mathbf{I} - \frac{\partial \mathbf{f}}{\partial \mathbf{y}}\right)^{-1}\left(\frac{\partial \mathbf{f}}{\partial \theta}\right),
\end{align}
where $\mathbf{I}$ is the identity matrix.
The inverse-matrix vector calculation in the equation above is performed using the BiCGSTAB algorithm \cite{van1992bicgstab}.
The gradient of this self-consistent iteration is implemented in xitorch \cite{kasim2020xitorch}.

\subsection{Gaussian basis}

The differentiable KS-DFT employs a contracted Gaussian basis.
The calculation of the Gaussian basis in 3D space is performed using libcgto, a library within PySCF \cite{sun2018pyscf}.
Libcgto is a library that can generate code to compute a Gaussian basis of any order.
As libcgto does not provide an automatic differentiation feature, it has to be wrapped by PyTorch to enable the automatic differentiation mode through the integrals calculated by libcgto.

A single Gaussian basis centered at $\mathbf{r_0}$ of order $l, m, n$ is expressed as
\begin{equation}
    g_{lmn}(\mathbf{r}; \alpha, c, \mathbf{r_0}) = c(x - x_0)^l (y - y_0)^m (z - z_0)^n e^{-\alpha ||\mathbf{r} - \mathbf{r_0}||^2},
\end{equation}
where $c$ is the basis coefficient and $\alpha$ is the exponential parameter.
The contracted Gaussian basis is just a linear combination of the single Gaussian bases above.

The gradient back-propagation calculation with respect to the parameters, ($c$, $\alpha$, and $\mathbf{r_0}$) can be expressed as
\begin{align}
    \frac{\partial \mathcal{L}}{\partial c} &= \frac{\partial \mathcal{L}}{\partial g_{lmn}} \frac{\partial g_{lmn}}{\partial c} \\
    \frac{\partial \mathcal{L}}{\partial \alpha} &= \frac{\partial \mathcal{L}}{\partial g_{lmn}}\frac{\partial g_{lmn}}{\partial \alpha} \\
    \frac{\partial \mathcal{L}}{\partial \mathbf{r_0}} &= \frac{\partial \mathcal{L}}{\partial g_{lmn}}\frac{\partial g_{lmn}}{\partial \mathbf{r_0}},
\end{align}
where 
\begin{align}
    \frac{\partial g_{lmn}}{\partial c} &= \frac{g_{lmn}}{c} \\
    \frac{\partial g_{lmn}}{\partial \alpha} &= -\left(g_{(l+2)mn} + g_{l(m+2)n} + g_{lm(n+2)}\right) \\
    \frac{\partial g_{lmn}}{\partial \mathbf{r_0}} &= -\nabla g_{lmn}.
\end{align}
The parameters required for the calculations above (e.g. the gradient of the basis and higher order bases) can be provided by libcgto from PySCF~\cite{sun2018pyscf}.

\subsection{Gaussian integrals}

Constructing the Fock matrix and the overlap matrix in Roothaan's equation requires the evaluation of Gaussian integrals.
The integrals typically involve 2-4 Gaussian bases of any order.
Those integrals are:
\begin{align}
    S_{ijk-lmn} &= \int g_{ijk}(\mathbf{r})g_{lmn}(\mathbf{r})\ \mathrm{d}\mathbf{r} \\
    K_{ijk-lmn} &= -\frac{1}{2} \int g_{ijk}(\mathbf{r})\nabla^2 g_{lmn}(\mathbf{r})\ \mathrm{d}\mathbf{r} \\
    C_{ijk-lmn} &= \sum_c \int g_{ijk}(\mathbf{r})\frac{Z_c}{|\mathbf{r} - \mathbf{r_c}|}g_{lmn}(\mathbf{r})\ \mathrm{d}\mathbf{r} \\
    E_{ijk-lmn-pqr-stu} &= \nonumber \\
    \int g_{ijk}(\mathbf{r}) g_{lmn}(\mathbf{r})&\frac{1}{|\mathbf{r}-\mathbf{r'}|}g_{pqr}(\mathbf{r'}) g_{stu}(\mathbf{r'})\ \mathrm{d}\mathbf{r}\mathrm{d}\mathbf{r'},
\end{align}
where $S$ is the overlap term, $K$ is the kinetic term, $C$ is the ionic Coulomb term of an ion of charge $Z_c$ at coordinate $\mathbf{r_c}$, and $E$ is the two-electron Coulomb term.
These integrals can be calculated efficiently by libcint \cite{sun2015libcint}.

The gradients of a loss function $\mathcal{L}$ with respect to the basis parameters $\theta = \{c, \alpha, \mathbf{r_0}\}$ are:
\begin{align}
    \frac{\partial \mathcal{L}}{\partial \theta} =& \left(\frac{\partial \mathcal{L}}{\partial \theta}\right)_S + \left(\frac{\partial \mathcal{L}}{\partial \theta}\right)_K + \left(\frac{\partial \mathcal{L}}{\partial \theta}\right)_C + \left(\frac{\partial \mathcal{L}}{\partial \theta}\right)_E\\
    \left(\frac{\partial \mathcal{L}}{\partial \theta}\right)_S =& \frac{\partial \mathcal{L}}{\partial S} \left[ \int \frac{\partial g_{ijk}(\mathbf{r})}{\partial \theta}g_{lmn}(\mathbf{r})\ \mathrm{d}\mathbf{r} \right. \nonumber \\
    &+ \left. \int g_{ijk} \frac{\partial g_{lmn}(\mathbf{r})}{\partial \theta}\ \mathrm{d}\mathbf{r} \right] \\
    \left(\frac{\partial \mathcal{L}}{\partial \theta}\right)_K =& -\frac{1}{2}\frac{\partial \mathcal{L}}{\partial K} \left[ \int \frac{\partial g_{ijk}(\mathbf{r})}{\partial \theta}\nabla^2 g_{lmn}(\mathbf{r})\ \mathrm{d}\mathbf{r}\right. \nonumber \\
    &+ \left. \int g_{ijk} \nabla^2 \frac{\partial g_{lmn}(\mathbf{r})}{\partial \theta}\ \mathrm{d}\mathbf{r}\right] \\
    \left(\frac{\partial \mathcal{L}}{\partial \theta}\right)_C =& \frac{\partial \mathcal{L}}{\partial C} \sum_c \left[ \int \frac{\partial g_{ijk}(\mathbf{r})}{\partial \theta}\frac{Z_c}{|\mathbf{r} - \mathbf{r_c}|} g_{lmn}(\mathbf{r})\ \mathrm{d}\mathbf{r} \right. \nonumber \\
    &+ \left. \int g_{ijk} \frac{Z_c}{|\mathbf{r} - \mathbf{r_c}|} \frac{\partial g_{lmn}(\mathbf{r})}{\partial \theta}\ \mathrm{d}\mathbf{r} \right] \\
    \left(\frac{\partial \mathcal{L}}{\partial \theta}\right)_E =& \frac{\partial \mathcal{L}}{\partial E} \left[ \int \frac{\partial g_{ijk}(\mathbf{r})}{\partial \theta}\frac{g_{lmn}(\mathbf{r}) g_{pqr}(\mathbf{r'}) g_{stu}(\mathbf{r'})}{|\mathbf{r} - \mathbf{r'}|} \ \mathrm{d}\mathbf{r} \right. \nonumber \\
    &+ \int \frac{\partial g_{lmn}(\mathbf{r})}{\partial \theta}\frac{g_{ijk}(\mathbf{r}) g_{pqr}(\mathbf{r'}) g_{stu}(\mathbf{r'})}{|\mathbf{r} - \mathbf{r'}|} \ \mathrm{d}\mathbf{r} \nonumber \\
    &+ \int \frac{\partial g_{pqr}(\mathbf{r'})}{\partial \theta}\frac{g_{ijk}(\mathbf{r}) g_{lmn}(\mathbf{r}) g_{stu}(\mathbf{r'})}{|\mathbf{r} - \mathbf{r'}|} \ \mathrm{d}\mathbf{r} \nonumber \\
    &+ \left. \int \frac{\partial g_{stu}(\mathbf{r'})}{\partial \theta}\frac{g_{ijk}(\mathbf{r}) g_{lmn}(\mathbf{r}) g_{pqr}(\mathbf{r'})}{|\mathbf{r} - \mathbf{r'}|} \ \mathrm{d}\mathbf{r} \right].
\end{align}
If the parameter $\theta$ does not belong to the basis $g_{\cdot}$, then $\partial g_{\cdot}/\partial \theta$ is 0.
Otherwise, the value of $\partial g_{\cdot}/\partial \theta$ for various parameters is listed in the previous subsection.
For the gradient with respect to the atomic position, $\mathbf{r_c}$, the expression is given by
\begin{align}
    \frac{\partial \mathcal{L}}{\partial \mathbf{r_c}} =& \frac{\partial \mathcal{L}}{\partial C} \left[\int \nabla g_{ijk}(\mathbf{r}) \frac{Z_c}{|\mathbf{r} - \mathbf{r_c}|} g_{lmn}(\mathbf{r})\ \mathrm{d}\mathbf{r}\right. \\
    &+ \left. \int g_{ijk}(\mathbf{r}) \frac{Z_c}{|\mathbf{r} - \mathbf{r_c}|} \nabla g_{lmn}(\mathbf{r})\ \mathrm{d}\mathbf{r}\right].
\end{align}
The expression above is obtained by performing partial integration.
As the derivative of the Gaussian basis with respect to its parameters is also a Gaussian basis, the required integrals can be provided by libcint \cite{sun2015libcint}.
Although the basis and position optimization was not performed during the xc neural network training, we present these gradients for the sake of completeness.

\subsection{Exchange-correlation energy and potential}
The exchange-correlation (xc) energy in our calculation is a hybrid between available xc energy functionals (the LDA \cite{kohn1965self-ks} and PBE \cite{perdew1996generalized-gga-pbe} models) and a neural network xc energy.
The gradient back-propagation through the xc neural network is easily provided by PyTorch.
However, as we use Libxc \cite{lehtola2018recent} v6.0.0 to calculate the previously available xc energy, it has to be wrapped by PyTorch to provide the gradient back-propagation calculation.
As Libxc can also calculate the gradient of the energy with respect to its parameters, providing the automatic differentiation feature by wrapping it with PyTorch is just a matter of calling the right function in Libxc.

If the PyTorch wrapper for Libxc is provided, calculating the xc potential can be done using the automatic differentiation feature provided by PyTorch.
The component of the Fock matrix from the xc potential for the LDA spin-polarized case is
\begin{equation}
    V^{\rm (LDA)}_{ij} = \int \phi_i^*(\mathbf{r}) v^{\rm(LDA)}_s(\mathbf{r}) \phi_j(\mathbf{r})\ \mathrm{d}\mathbf{r},
\end{equation}
where the subscript $s$ can be $\uparrow$ or $\downarrow$ representing the spin of the electron, and $v_s^{\rm (LDA)}$ is the xc potential,
\begin{equation}
    v^{\rm (LDA)}_s = \frac{\partial}{\partial n_s}\int n \varepsilon^{\rm(LDA)}(n_\uparrow, n_\downarrow)\ \mathrm{d}\mathbf{r},
\end{equation}
with $\varepsilon^{\rm(LDA)}$ the energy density per unit volume, per electron, provided by Libxc.
The 3D spatial integration is performed using a SG-3 integration grid \cite{dasgupta2017standard-sg3} with the Becke scheme to combine multi-atomic grids \cite{becke1988multicenter}.

For spin-polarized GGA, the potential linear operator is \cite{martin2020electronic}
\begin{align}
    V^{\rm(GGA)}_{ij} &= \int \phi_i^*(\mathbf{r}) v^{\rm(GGA)}_s(\mathbf{r}) \phi_j(\mathbf{r})\ \mathrm{d}\mathbf{r}\\
    v_{\rm GGAs} =& \frac{\partial}{\partial n_s}\int n \varepsilon^{\rm(GGA)}(n_\uparrow, n_\downarrow, \nabla n_\uparrow, \nabla n_\downarrow)\ \mathrm{d}\mathbf{r}\ + \nonumber \\ 
    &\left(\frac{\partial}{\partial \nabla n_s}\int n \varepsilon^{\rm(GGA)}\ \mathrm{d}\mathbf{r}\right)\cdot \nabla,
\end{align}
where the derivatives with respect to $n_s$ and $\nabla n_s$ can be performed by automatic differentiation, and the last $\nabla$ term is applied to the basis which can be provided by libcgto from PySCF \cite{sun2018pyscf}.

\subsection{(Near-)degenerate case of eigendecomposition}

One challenge in propagating the gradient backward for some molecules is the degeneracy of eigenvalues in the eigendecomposition.
The eigendecomposition of a real and symmetric matrix, $\mathbf{A}$, can be written as
\begin{equation}
    \mathbf{Av}_i = \lambda_i\mathbf{v}_i
\end{equation}
with $\lambda_i$ and $\mathbf{v}_i$ the $i$-th eigenvalue and eigenvector, respectively.

In the non-degenerate case, the gradient of a loss function $\mathcal{L}$ with respect to each element of matrix $\mathbf{A}$ can be written as
\begin{equation}
    \frac{\partial\mathcal{L}}{\partial \mathbf{A}} = -\sum_j\sum_{i\neq j} (\lambda_i - \lambda_j)^{-1}\mathbf{v}_i\mathbf{v}_i^T\frac{\partial\mathcal{L}}{\partial \mathbf{v}_j}\mathbf{v}_j^T.
\end{equation}
A problem arises when degeneracy appears as a term of $0^{-1}$ appears in the equation above.

To solve this problem, we follow the equation from Kasim~\cite{kasim2020derivatives-degen}, which can be written simply as
\begin{equation}
    \frac{\partial\mathcal{L}}{\partial \mathbf{A}} = -\sum_j\sum_{i\neq j, \lambda_i\neq \lambda_j} (\lambda_i - \lambda_j)^{-1}\mathbf{v}_i\mathbf{v}_i^T\frac{\partial\mathcal{L}}{\partial \mathbf{v}_j}\mathbf{v}_j^T.
\end{equation}
The equation above is only valid as long as the loss function does not depend on which linear combinations of the degenerate eigenvectors are used and the elements in matrix $\mathbf{A}$ depends on some parameters that always make the matrix $\mathbf{A}$ real and symmetric.
For more detailed derivation, please see~\cite{kasim2020derivatives-degen}.

\section{Neural network training}

\subsection{Dataset}

The atomization energy dataset is compiled from NIST CCCBDB \cite{NIST_CCCBDB}.
CCCBDB contains the experimental data for enthalpy of atomization energy.
To convert it into electronic atomization energy which can be calculated by KS-DFT calculations, the zero-point vibration energy must be excluded from the enthalpy of atomization energy.
Some diatomic molecules have the experimental values of the zero-point energy in CCCBDB from \cite{irikura2007experimental-zpe}.
For the rest of the molecules, the zero-point energy is estimated from the fundamental vibrations of the molecules which are also available from CCCBDB.

The ionization potential dataset is compiled by getting the ionization energy of neutral atoms from NIST ASD~\cite{NIST_ASD}.

\subsection{Neural networks}

The neural network takes an input of the density, $n$, spin polarized density, $\xi=(n_\uparrow - n_\downarrow)/n$, and normalized density gradient, $s=|\nabla n|/(24\pi^2 n^4)^{1/3}$ (for \xcdnn{}-PBE).
As the density ($n$) and the normalized density gradient ($s$) can have values between 0 up to a very large number, these values are renormalized by applying $\log(1+x)$, i.e. $n\leftarrow \log(1+n)$ and $s\leftarrow \log(1+s)$.

\subsection{Training}

The parameters of \xcdnn{} are updated based on the gradient of a loss function with respect to its parameters.
The loss function is given by a combination of three terms
\begin{align}
    \mathcal{L} &= \frac{w_{ae}}{N_{ae}} \sum_{i=1}^{N_{ae}}\left(\hat{E}_i^{(ae)} - E_i^{(ae)}\right)^2 + \frac{w_{ip}}{N_{ip}} \sum_{i=1}^{N_{ip}}\left(\hat{E}_i^{(ip)} - E_i^{(ip)}\right)^2 \nonumber \\
    &+ \frac{w_{dp}}{N_{dp}} \sum_{i=1}^{N_{dp}}\int\left[\hat{n}_i(\mathbf{r}) - n_i(\mathbf{r})\right]^2\mathrm{d}\mathbf{r},
\end{align}
where $w_{(\cdot)}$ are the weights assigned to each type of the data, $N_{(\cdot)}$ are the number of entries of each datatype in the dataset, $E^{(ae)}$ and $E^{(ip)}$ are, respectively, the predicted atomization energy and ionization potential using \xcdnn{} and KS-DFT, $n$ is the electron density profile predicted with \xcdnn{} and KS-DFT.
The variables with a hat, $\hat{E}^{(\cdot)}$ and $\hat{n}$, are the target variables that were obtained from experimental databases (for energies) and CCSD calculations (for electron density profiles).
The density in the loss function is represented in atomic unit while the energy is in Hartree.

The \xcdnn{} is trained using the RAdam \cite{liu2019variance-radam} algorithm with a constant learning rate of $10^{-4}$.
The weights for validation loss are:
\begin{itemize}
    \item \xcdnn{}: $w_{ae}=1340$, $w_{dp}=170$, $w_{ip}=0$;
    \item \xcdnn{}-IP: $w_{ae}=1340$, $w_{dp}=170$, $w_{ip}=1340$.
\end{itemize}
Those values are chosen to make the validation loss using LDA are about 1 for each properties.
Other combination of weights are also acceptable, depending on what one would prioritize.
The weights for training the neural networks are then chosen manually to make the validation loss minimum.
Those weights are:
\begin{itemize}
    \item \xcdnn{}-LDA: $w_{ae}=1340$, $w_{dp}=170$, $w_{ip}=0$;
    \item \xcdnn{}-PBE: $w_{ae}=1340$, $w_{dp}=5360$, $w_{ip}=0$;
    \item \xcdnn{}-LDA-IP: $w_{ae}=1340$, $w_{dp}=170$, $w_{ip}=1340$;
    \item \xcdnn{}-PBE-IP: $w_{ae}=1340$, $w_{dp}=5360$, $w_{ip}=2680$.
\end{itemize}

\subsection{Density profile deviation}

The density profile deviation for DP 99 calculation is obtained by
\begin{equation}
    \mathcal{L}_{dp} = \int [\hat{n}(\mathbf{r}) - n(\mathbf{r})]^2\ \mathrm{d}\mathbf{r},
\end{equation}
where $\hat{n}(\mathbf{r})$ is the density profile calculated by CCSD and $n(\mathbf{r})$ is the density profile obtained using XCNN.

\section{Additional analysis}
\subsection{GGA functionals comparison}

To compare XCNN-PBE and XCNN-PBE-IP with other GGA functionals, we collected 48 GGA functionals from Libxc~\cite{lehtola2018recent} v6.0.0.
The GGA functionals considered here are the functionals that have both exchange and correlation parts in Libxc.
PySCF is used for the evaluation of functionals other than XCNNs.
The comparison results on the GGA functionals can be found on Table \ref{tab:ggas-results}.

\subsection{Density profile regularization effect}

\begin{figure}
    \centering
    \includegraphics[width=\linewidth]{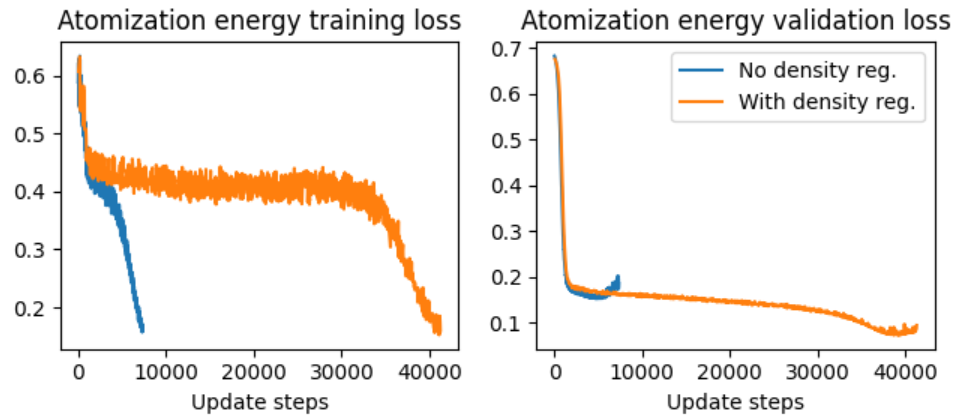}
    \caption{The training and validation losses for cases with no density profile regularization and with density profile regularization.}
    \label{fig:dens-vs-no-dens}
\end{figure}

As pointed out by ref.~\cite{nagai2020completing}, adding density profile in the loss function acts as a regularizer to avoid overfitting.
We observed a similar situation in our case.
As seen on Figure \ref{fig:dens-vs-no-dens}, the case with no density profile regularization overfits very quickly with relatively higher validation loss on atomization energy compared to the one with density regularization.

\subsection{Comparison with CCSD}

\begin{figure}[b]
    \centering
    \includegraphics[width=\linewidth]{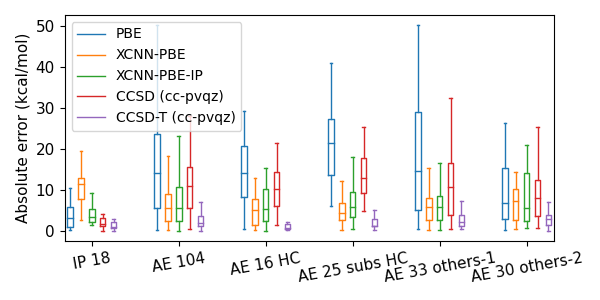}
    \caption{
    Statistical comparison of PBE, XCNN-PBE, CCSD, and CCSD(T) for each group.
    The line in the box represents the median of the absolute error.
    For clarity, outliers are not shown.}
    \label{fig:top-performer}
\end{figure}

The detailed comparison of XCNN-PBE and XCNN-PBE-IP with CCSD is shown in Figure \ref{fig:top-performer}.

\section{Molecules in the test dataset}

The 104 molecules in the test dataset, as well as the subset they belong, to are listed below.
\begin{enumerate}    \item LiH (Lithium Hydride): Others-1
    \item BeH (beryllium monohydride): Others-1
    \item CH (Methylidyne): HC
    \item CH$_3$ (Methyl radical): HC
    \item CH$_4$ (Methane): HC
    \item NH (Imidogen): Others-1
    \item NH$_2$ (Amino radical): Others-1
    \item NH$_3$ (Ammonia): Others-1
    \item OH (Hydroxyl radical): Others-1
    \item H$_2$O (Water): Others-1
    \item HF (Hydrogen fluoride): Others-1
    \item SiH$_2$ (silicon dihydride): Others-2
    \item SiH$_3$ (Silyl radical): Others-2
    \item SiH$_4$ (Silane): Others-2
    \item PH$_3$ (Phosphine): Others-2
    \item H$_2$S (Hydrogen sulfide): Others-2
    \item HCl (Hydrogen chloride): Others-2
    \item Li$_2$ (Lithium diatomic): Others-1
    \item LiF (lithium fluoride): Others-1
    \item C$_2$H$_2$ (Acetylene): HC
    \item C$_2$H$_4$ (Ethylene): HC
    \item C$_2$H$_6$ (Ethane): HC
    \item CN (Cyano radical): Others-1
    \item HCN (Hydrogen cyanide): Others-1
    \item CO (Carbon monoxide): Others-1
    \item HCO (Formyl radical): Others-1
    \item H$_2$CO (Formaldehyde): Others-1
    \item CH$_3$OH (Methyl alcohol): Subs HC
    \item N$_2$ (Nitrogen diatomic): Others-1
    \item N$_2$H$_4$ (Hydrazine): Others-1
    \item NO (Nitric oxide): Others-1
    \item O$_2$ (Oxygen diatomic): Others-1
    \item H$_2$O$_2$ (Hydrogen peroxide): Others-1
    \item F$_2$ (Fluorine diatomic): Others-1
    \item CO$_2$ (Carbon dioxide): Others-1
    \item Na$_2$ (Disodium): Others-2
    \item Si$_2$ (Silicon diatomic): Others-2
    \item P$_2$ (Phosphorus diatomic): Others-2
    \item S$_2$ (Sulfur diatomic): Others-2
    \item Cl$_2$ (Chlorine diatomic): Others-2
    \item NaCl (Sodium Chloride): Others-2
    \item SiO (Silicon monoxide): Others-2
    \item CS (carbon monosulfide): Others-2
    \item SO (Sulfur monoxide): Others-2
    \item ClO (Monochlorine monoxide): Others-2
    \item ClF (Chlorine monofluoride): Others-2
    \item Si$_2$H$_6$ (disilane): Others-2
    \item CH$_3$Cl (Methyl chloride): Subs HC
    \item HOCl (hypochlorous acid): Others-2
    \item SO$_2$ (Sulfur dioxide): Others-2
    \item BF$_3$ (Borane trifluoro-): Others-1
    \item BCl$_3$ (Borane trichloro-): Others-2
    \item AlF$_3$ (Aluminum trifluoride): Others-2
    \item AlCl$_3$ (Aluminum trichloride): Others-2
    \item CF$_4$ (Carbon tetrafluoride): Others-1
    \item CCl$_4$ (Carbon tetrachloride): Others-2
    \item OCS (Carbonyl sulfide): Others-2
    \item CS$_2$ (Carbon disulfide): Others-2
    \item CF$_2$O (Carbonic difluoride): Others-1
    \item SiF$_4$ (Silicon tetrafluoride): Others-2
    \item SiCl$_4$ (Silane tetrachloro-): Others-2
    \item N$_2$O (Nitrous oxide): Others-1
    \item ClNO (Nitrosyl chloride): Others-2
    \item NF$_3$ (Nitrogen trifluoride): Others-1
    \item O$_3$ (Ozone): Others-1
    \item F$_2$O (Difluorine monoxide): Others-1
    \item C$_2$F$_4$ (Tetrafluoroethylene): Others-1
    \item CF$_3$CN (Acetonitrile trifluoro-): Others-1
    \item CH$_3$CCH (propyne): HC
    \item CH$_2$CCH$_2$ (allene): HC
    \item C$_3$H$_4$ (cyclopropene): HC
    \item C$_3$H$_6$ (Cyclopropane): HC
    \item C$_3$H$_8$ (Propane): HC
    \item C$_4$H$_6$ (Methylenecyclopropane): HC
    \item C$_4$H$_6$ (Cyclobutene): HC
    \item CH$_3$CH(CH$_3$)CH$_3$ (Isobutane): HC
    \item C$_6$H$_6$ (Benzene): HC
    \item CH$_2$F$_2$ (Methane difluoro-): Subs HC
    \item CHF$_3$ (Methane trifluoro-): Subs HC
    \item CH$_2$Cl$_2$ (Methylene chloride): Subs HC
    \item CHCl$_3$ (Chloroform): Subs HC
    \item CH$_3$CN (Acetonitrile): Subs HC
    \item HCOOH (Formic acid): Subs HC
    \item CH$_3$CONH$_2$ (Acetamide): Subs HC
    \item C$_2$H$_5$N (Aziridine): Subs HC
    \item C$_2$N$_2$ (Cyanogen): Subs HC
    \item CH$_2$CO (Ketene): Subs HC
    \item C$_2$H$_4$O (Ethylene oxide): Subs HC
    \item C$_2$H$_2$O$_2$ (Ethanedial): Subs HC
    \item CH$_3$CH$_2$OH (Ethanol): Subs HC
    \item CH$_3$OCH$_3$ (Dimethyl ether): Subs HC
    \item C$_2$H$_4$O (Ethylene oxide): Subs HC
    \item CH$_2$CHF (Ethene fluoro-): Subs HC
    \item CH$_3$CH$_2$Cl (Ethyl chloride): Subs HC
    \item CH$_3$COF (Acetyl fluoride): Subs HC
    \item CH$_3$COCl (Acetyl Chloride): Subs HC
    \item C$_4$H$_4$O (Furan): Subs HC
    \item C$_4$H$_5$N (Pyrrole): Subs HC
    \item H$_2$ (Hydrogen diatomic): Others-1
    \item HS (Mercapto radical): Others-2
    \item C$_2$H (Ethynyl radical): HC
    \item CH$_3$O (Methoxy radical): Subs HC
    \item CH$_3$S (thiomethoxy): Subs HC
    \item NO$_2$ (Nitrogen dioxide): Others-1
\end{enumerate}

The 5 molecules excluded from the density profile deviation calculation are: CH (Methylidyne), OH (Hydroxyl radical), NO (Nitric oxide), ClO (Monochlorine monoxide), and HS (Mercapto radical).
Those molecules do not have a unique density profile when calculated using PySCF with PBE exchange correlation.

\newpage 
\begin{table*}
    \caption{Mean absolute error (MAE) in kcal/mol of the atomization energy and ionization potential for atoms and molecules in the test dataset.
    The column ``IP 18'' represents the deviation in ionization potential for atoms H-Ar.
    Column ``AE 104'' is the MAE of atomization energy of a collection of 104 molecules from ref.~\cite{curtiss1997assessment-g2}.
    Column ``DP 99'' is the difference of the density profile.
    Functionals above the line lie on the Pareto front for the deviations considered here.
    Some functionals that have very high errors typically do not converge on the self-consistent iterations.
    }
    \label{tab:ggas-results}
    \begin{ruledtabular}
    \begin{tabular}{rlrrr}
        No & Calculation & IP 18 & DP 99 ($\times 10^{-3}$) & AE 104 \\
        \hline
        \multicolumn{2}{l}{\textit{GGA functionals on the Pareto front}} & & \\
        1 & XCNN-PBE (this work) & 10.7 & 2.41 & 7.4 \\
        2 & XCNN-PBE-IP (this work) & 4.1 & 2.45 & 8.1 \\
        3 & PBE & 3.6 & 2.56 & 16.5 \\
        4 & RGE2 & 2.9 & 4.05 & 20.9 \\5 & XPBE & 3.4 & 2.70 & 9.4 \\
        6 & B97-D & 3.3 & 4.23 & 5.5 \\
        7 & EDF1 & 3.5 & 2.74 & 8.7 \\
        8 & HCTH 93 & 3.3 & 3.36 & 6.4 \\
        9 & OPBE-D & 4.4 & 2.49 & 8.0 \\
        10 & TH3 & 5.1 & 2.65 & 6.2 \\
        11 & TH4 & 10.9 & 3.05 & 5.5 \\
        \hline
        12 & AM05 & 3.6 & 13.08 & 37.0 \\
        13 & APBE & 4.0 & 2.86 & 11.1 \\
        14 & GAM & 3.9 & 15.96 & 7.8 \\
        15 & HCTH-A & 26.6 & 30.83 & 167.8 \\
        16 & N12 & 4.8 & 61.29 & 10.6 \\
        17 & PBE-MOL & 4.1 & 3.19 & 9.9 \\
        18 & PBE-SOL & 3.7 & 5.13 & 35.4 \\
        19 & PBEFE & 10.3 & 4.52 & 34.6 \\
        20 & PBEINT & 3.2 & 4.43 & 26.1 \\
        21 & PW91 & 4.3 & 2.90 & 17.1 \\
        22 & Q2D & 9.8 & 10.01 & 101.7 \\
        23 & SG4 & 4.8 & 3.37 & 17.5 \\
        24 & SOGGA11 & 7.6 & 7.23 & 10.1 \\
        25 & B97-GGA1 & 3.4 & 4.53 & 5.7 \\
        26 & BEEFVDW & 13.3 & 6.59 & 8.4 \\
        27 & HCTH-120 & 4.7 & 3.42 & 7.5 \\
        28 & HCTH-147 & 5.1 & 3.70 & 7.5 \\
        29 & HCTH-407 & 6.1 & 4.46 & 6.6 \\
        30 & HCTH-407P & 5.4 & 2.79 & 6.6 \\
        31 & HCTH-P14 & 18.6 & 7.61 & 12.6 \\
        32 & HCTH-P76 & 163.6 & 3668.60 & 324.9 \\
        33 & HLE16 & 37.6 & 71.94 & 23.3 \\
        34 & KT1 & 3.2 & 56.24 & 22.6 \\
        35 & KT2 & 4.6 & 87.23 & 10.6 \\
        36 & MOHLYP & 3.6 & 3.92 & 9.4 \\
        37 & MOHLYP2 & 9.5 & 4.52 & 67.6 \\
        38 & MPWLYP1W & 4.9 & 3.66 & 8.6 \\
        39 & OBLYP-D & 4.2 & 3.52 & 8.6 \\
        40 & OPWLYP-D & 4.2 & 3.53 & 8.5 \\
        41 & PBE1W & 5.5 & 2.75 & 13.8 \\
        42 & PBELYP1W & 5.5 & 3.41 & 11.3 \\
        43 & TH1 & nan & 13.18 & 288.2 \\
        44 & TH2 & 9.3 & 6.81 & 8.2 \\
        45 & TH-FC & 99.6 & 107.61 & 88.6 \\
        46 & TH-FCFO & 112.1 & 118.59 & 72.7 \\
        47 & TH-FCO & 110.7 & 150.56 & 60.2 \\
        48 & TH-FL & 111.0 & 97.92 & 26.2 \\
        49 & VV10 & 6.2 & 4.14 & 8.7 \\
        50 & XLYP & 4.6 & 3.74 & 7.8 \\
    \end{tabular}
    \end{ruledtabular}
\end{table*}

\newpage


%